# Magnetic behavior of Ni substituted $LiCoO_2$ – magnetization and electron paramagnetic resonance studies

Md. Mofasser Mallick and Satish Vitta*
Department of Metallurgical Engineering and Materials Science
Indian Institute of Technology Bombay
Mumbai 400076; India.

**Abstract**:

Single phase Ni substituted $LiCo_{1-x}Ni_xO_2$ solid solutions with $x \leq 0.15$ have been synthesized to study the effect of substitution on the magnetic behavior. Two different techniques, magnetic susceptibility and electron paramagnetic resonance (EPR) which provide information in two different time windows have been used. The solid solutions have been found to be single phase with large grains conforming to $R\bar{3}m$ rhombohedral structure. The lattice parameters 'c' and 'a' increase with increasing Ni substitution but with a nearly constant c/a ratio of ≈ 4.99 in all the cases indicating that the $CoO_6$ and $LiO_6$ octahedra do not undergo any Jahn-Teller distortions. The room temperature EPR absorption spectra clearly shows a peak in all the compounds at a field of 314 mT corresponding to a g-factor of 2.14. The peak width however is found to be a strong function of Ni substitution; increasing with increasing Ni from 3.6 mT to 6.5 mT for $x = 0.08$. The magnetization increases with decreasing temperature in all the compounds, a paramagnetic behavior, unlike Li-deficient compounds which show a Pauli paramagnetic susceptibility. Also, the magnetization exhibits thermal irreversibility which vanishes at large magnetic fields in all the compounds. The unsubstituted compound has discontinuities at 200 K and 50 K corresponding to magnetic transitions which disappear on substitution with 2 % Ni for Co. The effective magnetic moment $\mu_{eff}$ is found to increase from 0.32 $\mu_B$ to 0.66 $\mu_B$ on increasing the substitution to 0.15. The unique feature however is that all the compounds exhibit a clear magnetic hysteresis at room temperature with a finite coercivity. The coercivity increases from 31 Oe to 168 Oe for $x = 0.04$ and then decreases on further increasing of x. A deconvolution of the hysteresis loops clearly shows an increasing paramagnetic component with substitution.

* Email: Satish.vitta@iitb.ac.in

**Introduction:**

Oxides with a layered structure have strong correlations between charge, spin, orbital and lattice degrees of freedom and hence exhibit a wide variety of electronic phenomena varying from superconductivity[1-4] and ferromagnetism[5-8] to giant magnetoresistance.[9-11] One such compound is $LiCoO_2$ which is famously being used as an electrode material in solid state batteries.[12, 13] The stoichiometric variant of this compound is an insulator which becomes metallic on delithiation. The electrical conductivity has been found to change by 4 to 5 orders of magnitude on delithiation from 1.0 to ~ 0.7, reaching a maximum conductivity at this composition.[14] Delithiation is known to change the valence state of Co from 3+ to 4+ and result in hole doping in the triply degenerate $t_{2g}$ states of $Co^{3+}$.[14] The magnetic structure of stoichiometric $LiCoO_2$ and the delithiated compounds however is not clearly known and is a subject of intensive investigation by a variety of techniques.[15-17] In the stoichiometric compound, $CoO_6$ octahedra form a layer separating the Li-layers and the $Co^{3+}$ ions are known to have a two dimensional triangular lattice arrangement due to edge sharing octahedral.[18, 19] The $t^6_{2g}$ electronic configuration of $Co^{3+}$ ions results in a zero spin, S=0, non-magnetic state while the triangular arrangements results in magnetic frustration.[20] These factors are known to lead to a complex magnetic behavior in both the stoichiometric and delithiated compounds. The magnetic behavior in near stoichiometric compound, $Li_{1-x}CoO_2$ with $x \approx 0$ has been found to be due to the formation of an antiferromagnetic phase at low temperatures, < 65 K whose volume fraction was only ~ 20 %.[20, 21] The high temperature behavior however has been found to be due to a combination of charge disproportionation and spin state transition. The $Co^{3+}$ ions have been predicted to fluctuate between $Co^{2+}$ and $Co^{4+}$ states which have a higher spin state 1/2 compared to S = 0 for the diamagnetic $Co^{3+}$ state.[22, 23] This scenario is predicted to co-exist with spin state transition within the $Co^{3+}$ ions, i.e. $t^6_{2g}$ to $e_g'^4 a_g^1 e_g^1$ state with S = 1. Both these phenomena impart a net magnetic moment to the compound. These studies clearly indicate that the magnetic structure in $LiCoO_2$ is still not completely understood. Another compound belonging to this family is $LiNiO_2$ which has also been found to exhibit a variety of magnetic behaviors ranging from spin-glass like to antiferromagnetic with ferromagnetic in-layer exchange.[24-26] This compound has also been studied for application as electrode in batteries.

The presence of $Ni^{2+}$ ions at the 3a $Li^+$ site and exchange interactions between $Ni^{2+}$ and $Ni^{3+}$ in delithiated compounds is known to result in the formation of various magnetic structures in this compound.[27-30] The strength of interactions however is found to be small as the ordering of magnetic clusters leads to their breakdown and cannot be observed as fields of the order of 1000 Oe and above. These studies clearly show that the $ABO_2$ class of layered compounds exhibit a variety of electronic phenomena due to a complex interplay between the 4 degrees of freedom – charge, spin, orbital and lattice.

Solid solutions of $LiNiO_2$ and $LiCoO_2$ are also of importance as the substitution of transition metal ions in these compounds leads to changes in their electrochemical performance. Since Ni and Co have different valence states and hence different spin states, interchanging these ions leads to interesting magnetic behaviors.[31-35] Hence in the present work, solid solutions of the two layered $ABO_2$ compounds, $LiCoO_2$ and $LiNiO_2$, have been synthesized and their magnetic behavior has been investigated using complimentary techniques – susceptibility and magnetic hysteresis together with electron paramagnetic resonance. The amount of Ni has been varied up to 15 at. % of Co so that the compound always exists in a single phase state. The magnetic susceptibility has been studied as a function of temperature in the range 10 K to 300 K to understand the magnetic state in the substituted compound

**Experimental Methods:**

Single phase $LiCo_{1-x}Ni_xO_2$ compounds with $0 \leq x \leq 0.15$ where x is the atomic fraction have been synthesized using conventional solid state reaction method. High purity $LiCO_3$, $Co_2O_3$ and NiO were used as precursors. These were mixed in stoichiometric ratio in an agate mortar for 20 – 30 mins and subsequently calcined at 1073 K for 12 hours. After calcination, the resulting powder was again ground and annealed at 1123 K for 12 hours in ambient air atmosphere. The crystallographic structure and phases present in the powders was studied by X-ray diffraction using Cu-$K_\alpha$ radiation and subsequent Rietveld refinement of the powder diffraction patterns. The x-ray diffraction was performed with a rotating anode source and in a high resolution diffractometer with incident and diffracted beam monochromators. The variation of magnetization M with temperature T in zero field cooled (ZFC) and field-cooled (FC) conditions was studied in the range 10 K to 300 K in the presence of 1 kOe and 10 kOe external field. The magnetization hysteresis at 300 K has also been investigated to study the effect of dynamic magnetism at elevated temperatures. The electron paramagnetic resonance, EPR studies have been performed at room temperature using microwave radiation with an X-band frequency of ~ 9.4 GHz. The main advantage of using these techniques is that they cover very different time windows to observe the magnetic behavior. While the magnetic susceptibility studies provide static magnetic information the EPR technique provides dynamic information at ~ $10^{10}$ Hz frequency. X-ray photoelectron spectroscopy with Al $K_\alpha$ radiation was used to determine the valence state of cations present in all the compounds.

**Results and Discussion:**

The microstructure and grain size in the different compounds was studied by scanning electron microscopy and the electron micrographs are shown in Figure 1. The microstructure observed at two different locations shows that it is single phase with no secondary phases and that the structure is uniform. The grain size is found to be in the range 2 - 5 µm indicating formation of large, highly crystalline grains in all the compounds. The microstructure seen in Figure 1 is found to be typical and representative of the structure observed in all the compounds. The room temperature x-ray diffraction pattern of $LiCo_{1-x}Ni_xO_2$ with $0 \leq x \leq 0.15$ is shown in Figure 2 together with the results of structural refinement. The diffraction pattern in all the cases shows sharp peaks indicating the formation of a highly crystalline structure with large grains and no defects. The structural refinement was performed using the FullProf program for all the compositions and the fit parameters are given in Table 1. The diffraction pattern in all the cases could be indexed uniquely to $R\bar{3}m$ rhombohedral structure indicating the absence of other phases. The lattice parameters, c and a are found to increase with increasing Co substitution with Ni from 1.405 nm and 0.2815 nm for x = 0 to 1.4065 nm and 0.2820 nm for x = 0.15. The c/a ratio of the rhombohedral structure however is found to be nearly constant, ~ 4.99 indicating that this ratio plays a crucial role in stabilizing the rhombohedral structure.[36, 37] The proportionate increase of a and c to maintain a constant c/a ratio shows that the lattice expands iso-tropically with increasing Ni substitution for Co and retains regular geometry of the two octahedral – Co and Li without any Jahn-Teller type distortions. These results also show that the compounds are stoichiometric in nature as deviation from stoichiometry has been found to lead to Jahn-Teller type distortions to the Co and Li octahedra. The formation of a solid solution single phase and the variation of lattice parameters are in agreement with the lattice parameters variation found in $LiCoO_2$ – $LiNiO_2$ solid solutions synthesized by solid state reaction technique which show that both `a' and `c' increase with increasing Ni content.[34, 38] The physical properties of these compounds, electrochemical as well as magnetic are sensitive to the position of ions in the unit cell. In the case of $LiCoO_2$, Co-ions are known to occupy the 3b sites while the Li ions occupy the 3a sites.[37] In the $LiCo_{1-x}Ni_xO_2$ compounds the position of Ni ions is crucial for the magnetic properties. Hence to know the exact location of the different cations, Li, Co and Ni structural refinement was performed by inter-mixing their position and the results of best structural refinement are given in Table 1. These results show that while the Li and Co ions occupy their designated positions, i.e. 3a and 3b respectively, the position of Ni-ions could not be uniquely defined. Placing the Ni ions in Li sites instead of the Co sites did not alter significantly the goodness of fit showing that x-ray diffraction pattern is not sensitive to the exact position of Ni ions in the lattice. This is because of the absolute amount of substitution is small and also the atomic scattering factor of Ni and Co are nearly identical to resolve their position accurately. The oxidation state of Ni was determined by X-ray photoelectron spectroscopy and the results are shown in Figure 3. It is found that Ni exists in 2+ state in all the compounds. Earlier studies on determining the valence state of Ni in the solid solutions by electron energy-loss spectroscopy and x-ray absorption spectroscopy also revealed the presence of divalent Ni.[39, 40] All these studies clearly show that Ni is present in the divalent state in these compounds and not in the trivalent state. The presence of $Ni^{2+}$ with S = 1, a high spin state is important to understand the magnetic properties of the solid solutions. Hence based on these results, the structural formula for all the different compounds investigated can be written as $[Li^{1+} Ni^{2+}_{yx}]_{3b} [Ni^{2+}_{(1-y)x} Co_{1-x}]O_2$ with x ≤ 0.15 and y being the fraction of $Ni^{2+}$ in Li 3a sites, which could not be accurately determined.

The magnetic structure of nearly stoichiometric $LiCoO_2$ is not completely understood as $Co^{3+}$ ions can exist in the low spin, S = 0 state or the intermediate state with S = 1. If $Co^{3+}$ is in the diamagnetic low spin state, it should not exhibit any absorption in EPR. Hence to determine the presence of paramagnetic ions in the compounds EPR was performed at 300 K and the results are shown in Figure 4. A strong peak at a field of 314 mT corresponding to a spectroscopic g factor of 2.14±0.01 is observed in all the compounds.  A clear absorption peak at 300 K observed in earlier studies has been attributed to dynamic fluctuations of $Co^{3+}$ between low spin and intermediate spin states.[41] These results clearly show that all the compounds are magnetically active with unpaired electrons even at 300 K.  The width of these peaks however increases with increasing Ni-content from 3.6 mT for x = 0 to 6.5 mT for x = 0.08 and then decreases for x = 0.15. The peak-to-peak width of the absorption peak is inversely proportional to the spin-spin relaxation time[22] and it is found that the relaxation time in the present solid solutions varies between 4.9 x $10^{-10}$ S and 8.5 x $10^{-10}$ s in the different compounds, comparable to those observed for ions in inorganic compounds.[42]  These results clearly show the presence of magnetic ions in the compounds even at 300 K with an ordered arrangement.  The absolute magnetic behavior however changes with increasing Ni substitution – the lifetime of the paramagnetic entities decreases with increasing Ni-substitution.

In order to investigate the magnetic behavior further, the variation of magnetization, M with both temperature T and external magnetic field H have been studied. The variation of susceptibility χ with T in the presence of 1 kOe and 10 kOe external field in both zero-field cooled (ZFC) and field cooled (FC) states for all the different compounds is shown in Figure 5. The magnetization exhibits clear irreversibility between ZFC and FC at low fields, 1 kOe and low temperatures and this irreversibility vanishes at large fields in all the cases.  These results indicate the presence of magnetically ordered states such as superparamagnetic or spin-glass like in all the compounds.  The absolute value of M increases with increasing Ni indicating that magnetic entities as well as ordering of these entities increases in the compounds.  The x=0 compound, $LiCoO_2$ exhibits a discontinuous change in χ at 200 K – the magnetization increases monotonically till ~ 200 K at which temperature it drops sharply till ~ 185 K before increasing with decreasing T.  This behavior however vanishes completely for x = 0.02 Ni indicating that the magnetic transition at 200 K is limited to the pure compound alone with x = 0. Additionally, the magnetization increase with decreasing temperature for x = 0 compound changes sharply at ~ 50 K indicating a further magnetic transition to a more ordered phase.  A hump like feature in magnetic susceptibility of stoichiometric LiCoO$_2$ has been attributed earlier to the development of antiferromagnetic order, whose volume fraction however is only ~ 20 %, as determined by muon-spin rotation and relaxation.[20, 21] The variation of χ with T in compounds with x > 0 is typical of paramagnetic materials with a finite magnetic interaction. Magnetic fields > 1 kOe however result in suppressing the magnetic interactions and the system behaves as a paramagnet with increasing Ni content. The susceptibility χ (T) measured in FC mode in a field of 10 kOe of the substituted compounds with x > 0 has been fitted to Curie – Weiss law to determine the paramagnetic Curie temperature as well as the effective magnetic moment μ$_{eff}$ over the whole temperature range. The Curie-Weiss law is given by the relation;[43]

$$\chi (T) = N \mu^2_{eff} /3k_B(T-\Theta_p) + \chi_o \qquad (1)$$

where N is the no. of magnetic ions per mole in the compound, k$_B$ the Boltzmann constant, Θ$_p$ the Curie temperature and χ$_o$ the residual susceptibility. The fitting parameters are given in Table 2. The Curie temperature has been found to increase with increasing Ni and the effective magnetic moment is also found to increase with increasing Ni substitution, Figure 6. These results are in complete agreement with earlier reported values[44] and in the present case μ$_{eff}$ is found to increase from 0.32 μ$_B$ for x = 0 to 0.66 μ$_B$ for x = 0.15[45]. The temperature independent residual susceptibility in all the cases is found to be nearly same, ≈ 300x10$^{-6}$ emu mol$^{-1}$. This is in agreement with earlier studies wherein delithiation leading to effectively higher Co has been found to increase χ$_o$.[46]

The variation of M with H in the range ± 50 kOe at 300 K is shown in Figure 7 for all the different compounds. Several interesting features can be seen in Figure 7 – (i) M increases with increasing x, i.e. a behavior observed even in M-T results (ii) saturation tendency of M decreases with increasing x, i.e. as x increases M shows an increasing linearly increasing tendency with field H; and (iii) all the compounds have a clear coercivity Hc which increases till x = 0.04 and then decreases. The magnetization at 50 kOe increases from 0.06 emu g$^{-1}$ for x = 0 to 0.14 emug$^{-1}$ for x = 0.15 and the coercivity increases from 31 Oe to 168 Oe for x = 0.04 and then decreases to 53 Oe for x = 0.15. A stoichiometric LiCoO$_2$ well annealed in oxygen ambient has also been found to exhibit a large coercivity and a hysteretic magnetic behavior.[17] These results clearly show that all the compounds have a strong ferromagnetic component along with a paramagnetic component. The effect of substitution of Co with Ni without changing the stoichiometry of Li however has not been studied in detail earlier and hence is studied here in detail. Since the variation of M with field, Figure 7 clearly shows the presence of both paramagnetic and ferromagnetic components at room temperature it has been modeled using the following equation;[47]

$$M(H) = \left(2M_s^{ferro}/\pi\right) tan^{-1}\left[\frac{H \pm H_c}{H_c} tan(\pi S/2)\right] + \chi^{para} H \qquad (2)$$

where $M_s^{ferro}$ is the saturation magnetization of ferromagnetic phase, S the hysteresis loop squareness parameter and $\chi^{para}$ the paramagnetic susceptibility. The results of the fit are shown in Figure 7 together with the experimental data and the parameters obtained are given in Table 2. There are two points which require mention – the saturation magnetization of ferromagnetic component $M_s^{ferro}$ is nearly constant for all the different substitutions while the paramagnetic susceptibility increases with Ni substitution, Table 3. These results are in complete agreement with magnetization variation with temperature which show an increasing magnetization with increasing Ni and also an increasing paramagnetic nature, i.e. increasing linear component with increasing Ni.

Although a field dependent magnetic hysteresis behaviour has been observed earlier in the case of stoichiometric LiCoO2 in both as prepared and annealed conditions, exact reasons for this magnetic behaviour are not known and the complex magnetic behaviours are still intensely investigated and discussed. The ferromagnetic hysteresis behaviour in the present work is plausibly due to the presence of $Co^{3+}$ in the intermediate spin state and strong exchange interactions between $Co^{3+}$ ions. A strong EPR absorption peak observed at 300 K in the unsubstituted compound strongly supports this scenario. The $LiNiO_2$ compound on the other hand is known to be ferromagnetic due to the presence of low spin, S = 1/2 $Ni^{3+}$ ions. Any variation in stoichiometry in this compound has been found to change the magnetic behaviour significantly depending on either presence or absence of Ni ions in Li sites. The Ni ions in Li sites are found to be divalent in Li deficient compounds and this leads to antiferromagnetic inter-layer and ferromagnetic intra-layer exchange interactions. This complexity is amplified if Ni is substituted for Co in the LiCoO$_2$ compound. The substitution of Co with Ni in the LiCo$_{1-x}$Ni$_x$O$_2$ compounds in the present work is found to result in the formation of divalent Ni with a high spin, S = 1 magnetic state. This coupled with the presence of these ions in either Li sites or Co sites leads to variations in the magnetic behaviour. A strong EPR absorption peak at 300 K in all the substituted compounds, thermal irreversibility of magnetisation at low fields and a hysteric magnetisation at room temperature, all lead to the conclusion that the divalent Ni ions have a strong ferromagnetic character which is of predominantly intra-layer in nature. Additionally, in LiNi$_{1-y}$Co$_y$O$_2$ compounds, Co$^{3+}$ ions have been found to have a clustering tendency. Presence of such highly localized clusters can also alter the magnetic behaviour significantly.

**Conclusions:**

LiCoO$_2$ is a well-known electrode material used in solid state batteries. The magnetic behaviour of this compound and its non-stoichiometric variants however is still not completely understood. The magnetic behaviour of this compound is due to trivalent Co$^{3+}$ ions which have a tendency to exist in different states depending on the environment they are in. For e.g. it can undergo charge disproportionation into Co$^{2+}$ and Co$^{4+}$ which have entirely different magnetic signatures compared to Co$^{3+}$, depending on the amount of Li$^+$ ions in the compound so as to maintain a charge balance. The diamagnetic Co$^{3+}$ ions themselves can also exist in non-zero magnetic state due to spin dynamics. The present work is an attempt to understand the effect of substituting Co with Ni on the above mentioned magnetic behaviour of LiCoO$_2$. Small amounts of Co, < 0.15 has been substituted with Ni so that the compound exists in its native layered structure R$\overline{3}$m without undergoing any dissociation into secondary phases. Room temperature electron paramagnetic resonance clearly shows the presence of magnetically active ions in all the compounds with a finite spin-spin relaxation time. The static magnetization studies, both M-T and M-H clearly show that ferromagnetic and paramagnetic entities coexist. The fraction of ferromagnetic entities however decrease with increasing Ni substitution and the compounds become more and more paramagnetic in nature. The effective magnetic moment increases with increasing Ni content. The ferromagnetic coercivity increases before decreasing for substitutions > 0.04, clearly supporting the observation that paramagnetic component increases in the compounds. The valence state of Ni is found to be 2, a $d^8$ electronic configuration with S = 1. The exact location of these divalent Ni-ions, Li-planes or Co-planes, is not very clear. An identification of the location of these ions will reveal the extent of magnetic interactions, ferromagnetic vis-à-vis antiferromagnetic that exist in these compounds. Apart from the actual magnetic behaviour, this can have a significant influence on the reversibility of Li-ions intercalation process.

**Acknowledgements:**

The authors wish to acknowledge the Nanomission, Department of Science and Technology, Govt. of India for the financial support and IIT Bombay-Central facilities, SVSM and PPSM for magnetic characterization.

Table 1. The crystal structure parameters obtained by Rietveld refinement of x-ray diffraction pattern in all the compounds. 3a, 3b and 6c represent 3 different sites occupied by the 3 ions in R$\bar{3}$m crystal structure.

| Sl. No. | Chemical Composition | Lattice parameters | | 3b | 3a | 6c | Occupation fraction (%) | | | |
|---|---|---|---|---|---|---|---|---|---|---|
| | | a, nm | c, nm | | | | Li | Co | Ni | O |
| 1 | LiCoO$_2$ | 0.2815 | 1.4049 | Co | Li | O | 100 | 100 | 0 | 50 |
| 2 | LiCo$_{0.98}$Ni$_{0.02}$O$_2$ | 0.2816 | 1.4049 | Co, Ni | Li | O | 100 | 98 | 2 | 50 |
| 3 | LiCo$_{0.96}$Ni$_{0.04}$O$_2$ | 0.2816 | 1.4051 | Co, Ni | Li | O | 100 | 96 | 4 | 50 |
| 4 | LiCo$_{0.92}$Ni$_{0.08}$O$_2$ | 0.2817 | 1.4052 | Co, Ni | Li | O | 100 | 92 | 8 | 50 |
| 5 | LiCo$_{0.85}$Ni$_{0.15}$O$_2$ | 0.2820 | 1.4065 | Co, Ni | Li | O | 100 | 85 | 15 | 50 |

Table 2. The magnetic susceptibility is fit to Curie-Weiss law and the fitting parameters are given here. C = Curie-Weiss constant given by $\left(N\mu_{eff}^2/3k_B\right)$ where N is the no. of magnetic ions per mole, $\mu_{eff}$ the effective paramagnetic moment and $k_B$ the Boltzmann constant; $\chi_0$ the residual high temperature susceptibility in emu mol$^{-1}$Oe$^{-1}$; $\Theta_p$ the Curie temperature in K.

| x | 10 kOe (ZFC) | | | | 10 kOe (FC) | | | |
|---|---|---|---|---|---|---|---|---|
| | C | $\Theta_p$ | $\chi_0$ | $\mu_{eff}$ | C | $\Theta_p$ | $\chi_0$ | $\mu_{eff}$ |
| 0.02 | 0.012 | -0.16 | .0003 | .31 | 0.013 | -0.3 | .0003 | .32 |
| 0.04 | 0.015 | -2.8 | .0004 | .35 | 0.015 | -3.1 | .0004 | .35 |
| 0.08 | 0.021 | 0.23 | .0003 | .41 | 0.021 | 0.16 | .0003 | .41 |
| 0.15 | 0.054 | 3.04 | .0003 | .66 | 0.054 | 3.22 | .0003 | .66 |

Table 3. The total magnetization M is due to a combination of two phases, ferromagnetic and paramagnetic and is modeled using eq. (2) in the text. The best fit parameters are given.

| Ni substitution, x | $M_S^{ferro}$, emu g$^{-1}$ | $H_c$, Oe | $\chi^{para}$, emu g$^{-1}$Oe$^{-1}$ |
|---|---|---|---|
| 0 | 0.0331 | 31.5 | 5.44×10$^{-7}$ |
| 0.02 | 0.0273 | 78.63 | 9.51×10$^{-7}$ |
| 0.04 | 0.037 | 168.5 | 9.98×10$^{-7}$ |
| 0.08 | 0.0278 | 115.2 | 1.53×10$^{-6}$ |
| 0.15 | 0.0334 | 52.8 | 2.16×10$^{-6}$ |

**Figure Captions:**

Figure 1. Scanning electron micrographs show presence of large, defect free grains in all the different $LiCo_{1-x}Ni_xO_2$ compounds. The microstructure shows a single phase with no secondary phases. Typical microstructure in x = 0 (a) and x = 0.04 compounds.

Figure 2. X-ray diffraction patterns together with Rietveld refinement of the diffraction patterns shows a single phase with $R\bar{3}m$ rhombohedra, structure (a) in all the five compounds. The lattice parameters a and c increase with increasing x in $LiCo_{1-x}Ni_xO_2$ (b) while the c/a ration remains nearly constant, inset in (b).

Figure 3. X-ray absorption spectra from x = 0, 0.04 and 0.15 compounds shows Co in trivalent state while Ni is in divalent state in $LiCo_{1-x}Ni_xO_2$.

Figure 4. Room temperature electron paramagnetic resonance spectrum in all the five compounds with different amounts of Ni has a clear absorption peak at 314 mT field corresponding to a g-factor of 2.14. The inset shows the actual absorption peak in all the compounds.

Figure 5. The variation of magnetic susceptibility $\chi$ with temperature T in an external field of 1 kOe (a) and 10 kOe (b). The field cooled (FC) and zero-field cooled (ZFC) susceptibility shows irreversibility at low fields and it vanishes at large fields. The paramagnetic susceptibility fitted to Curie-Weiss law is shown by the continuous line in (b).

Figure 6. The effective paramagnetic moment $\mu_{eff}$ obtained by fitting the susceptibility to Curie-Weiss law increases with increasing Co substitution with Ni in $LiCo_{1-x}Ni_xO_2$.

Figure 7. The magnetic hysteresis loops at room temperature in fields up to 50 kOe show increasing magnetization with Ni substitution and an increasing paramagnetic component. The line through the data is a fit to the coexistence of two magnetic phases, ferromagnetic and paramagnetic model, eq. (2) in the text. Inset is the magnetization in small fields clearly showing coercivity in all the $LiCo_{1-x}Ni_xO_2$ compounds.

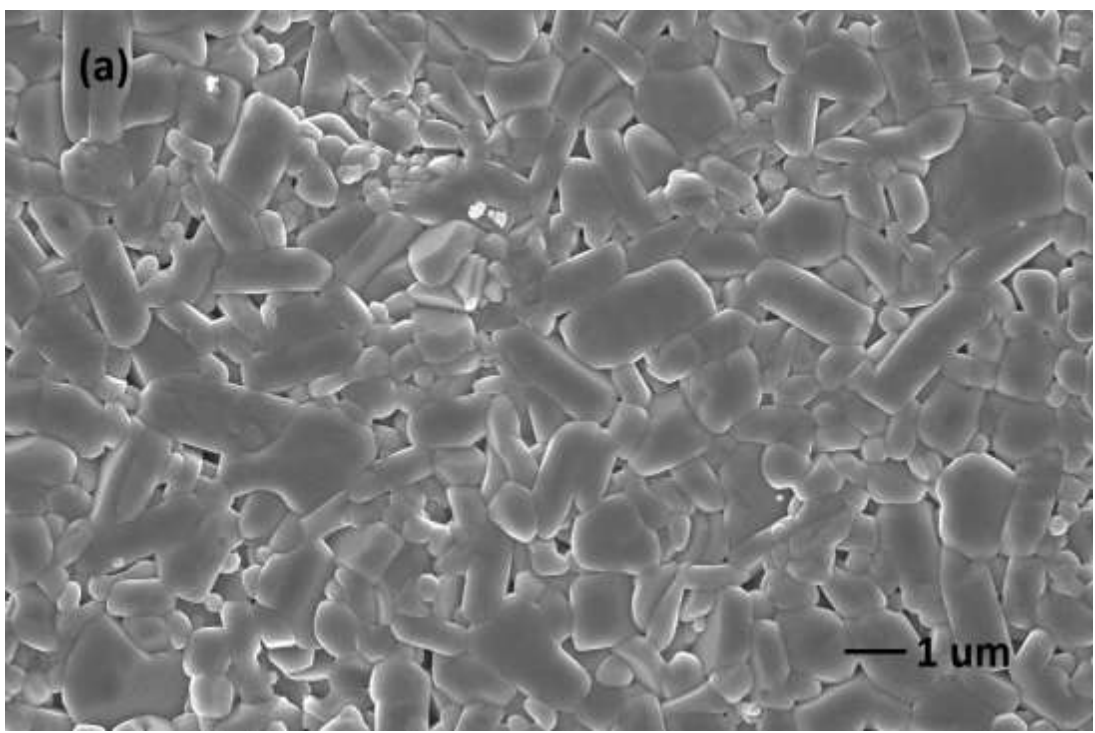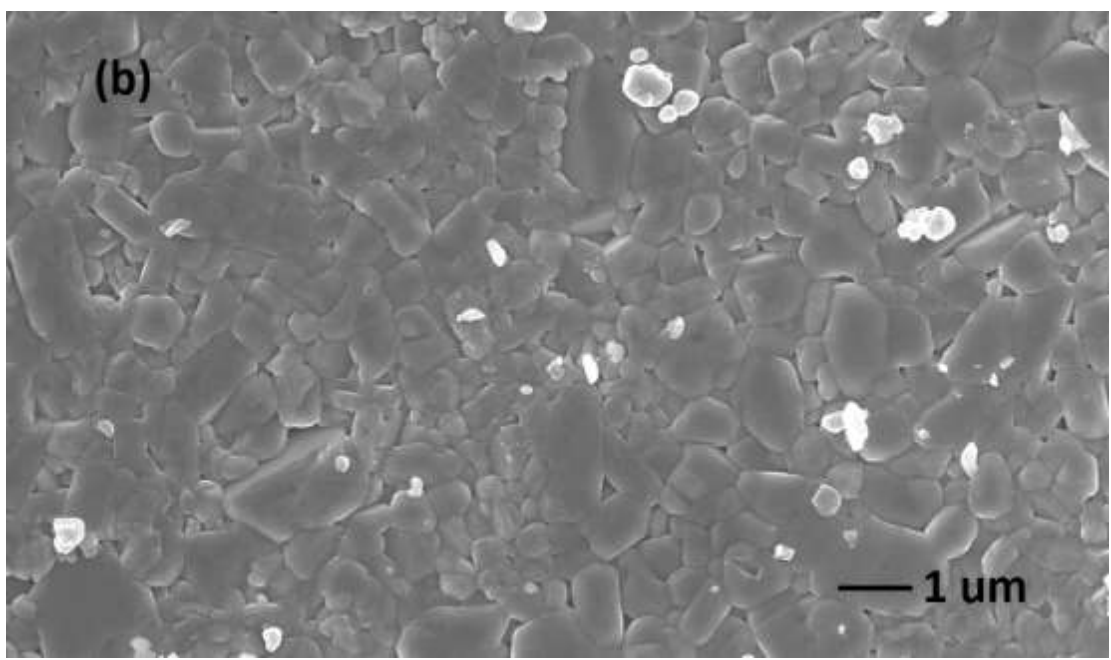

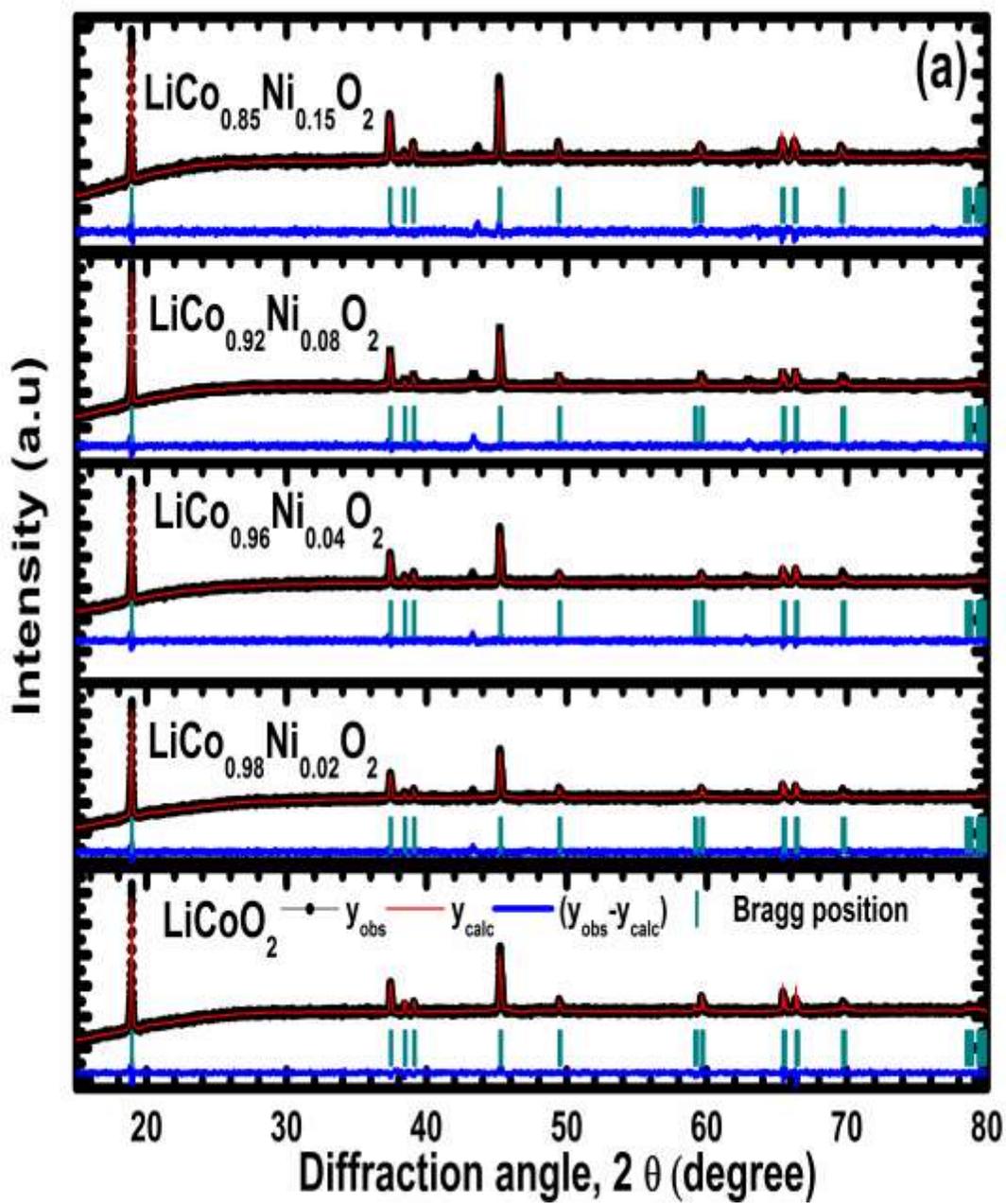

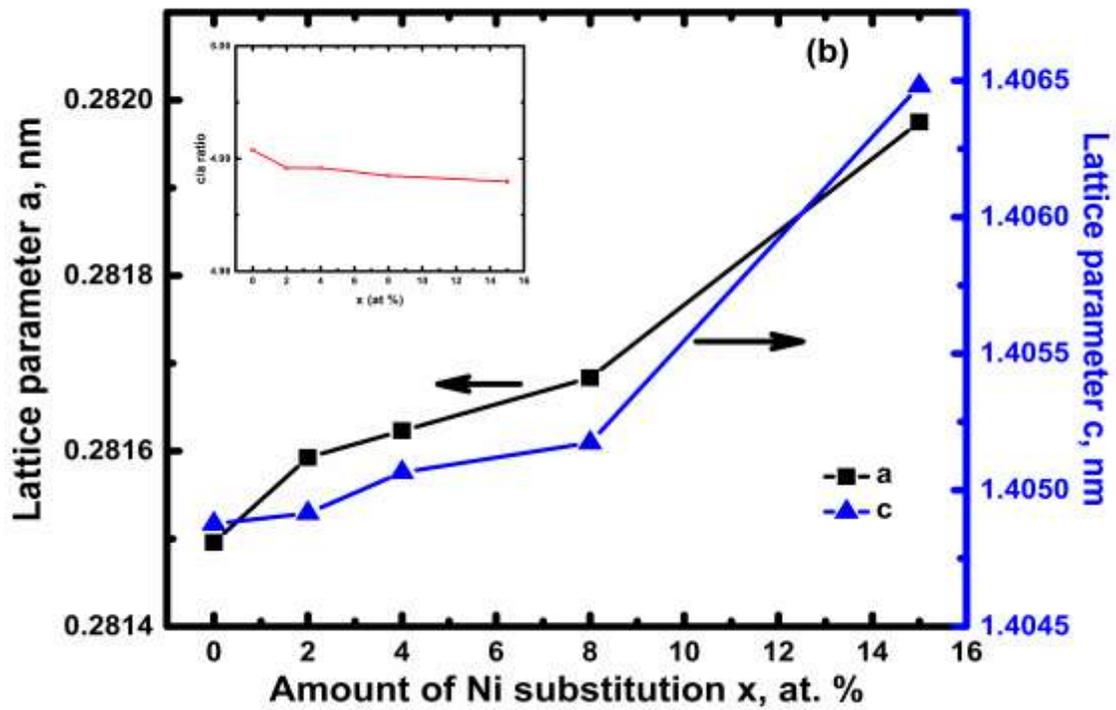

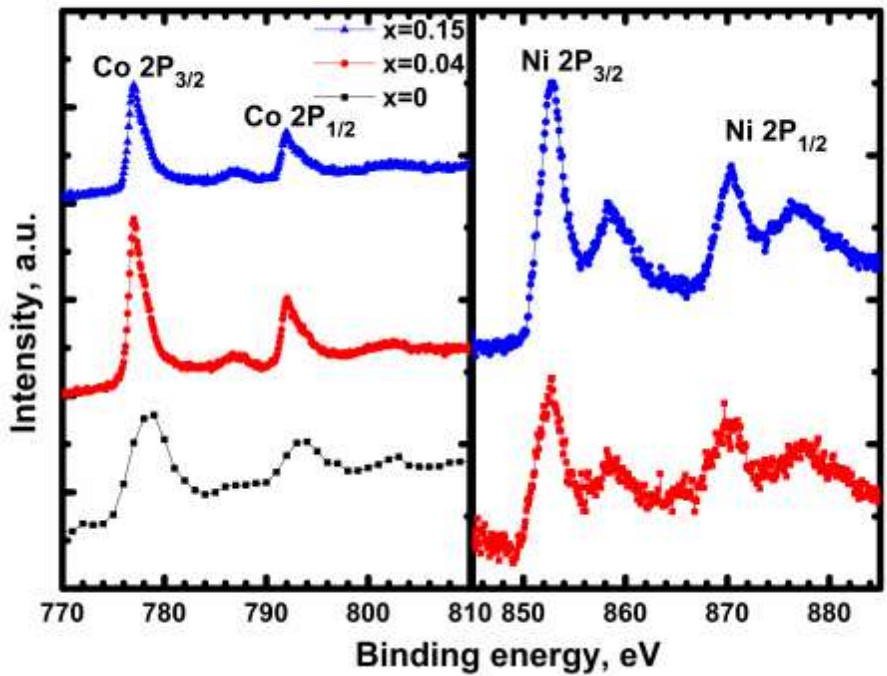

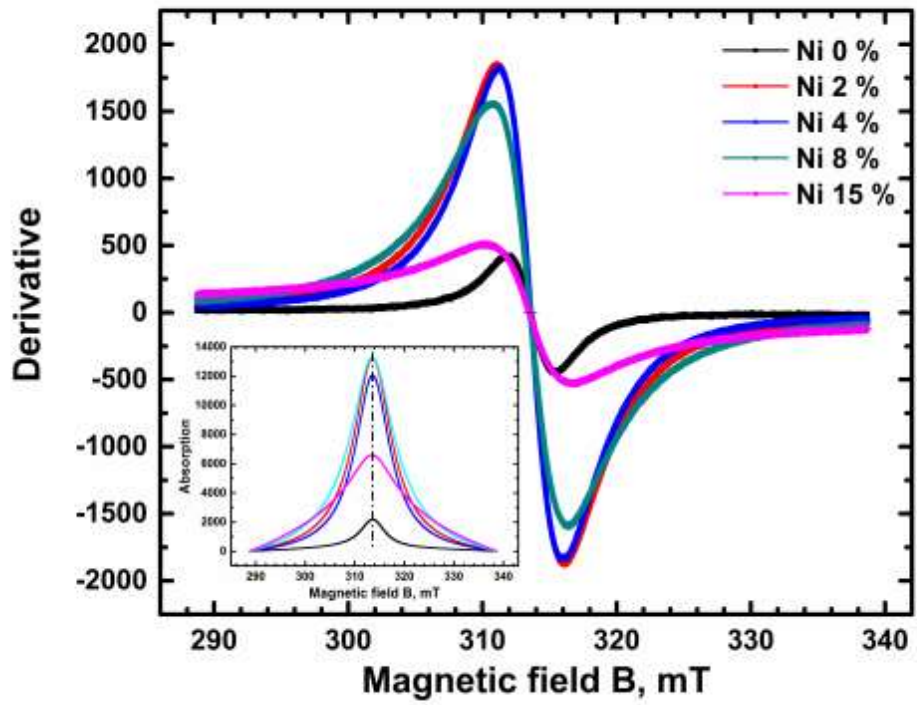

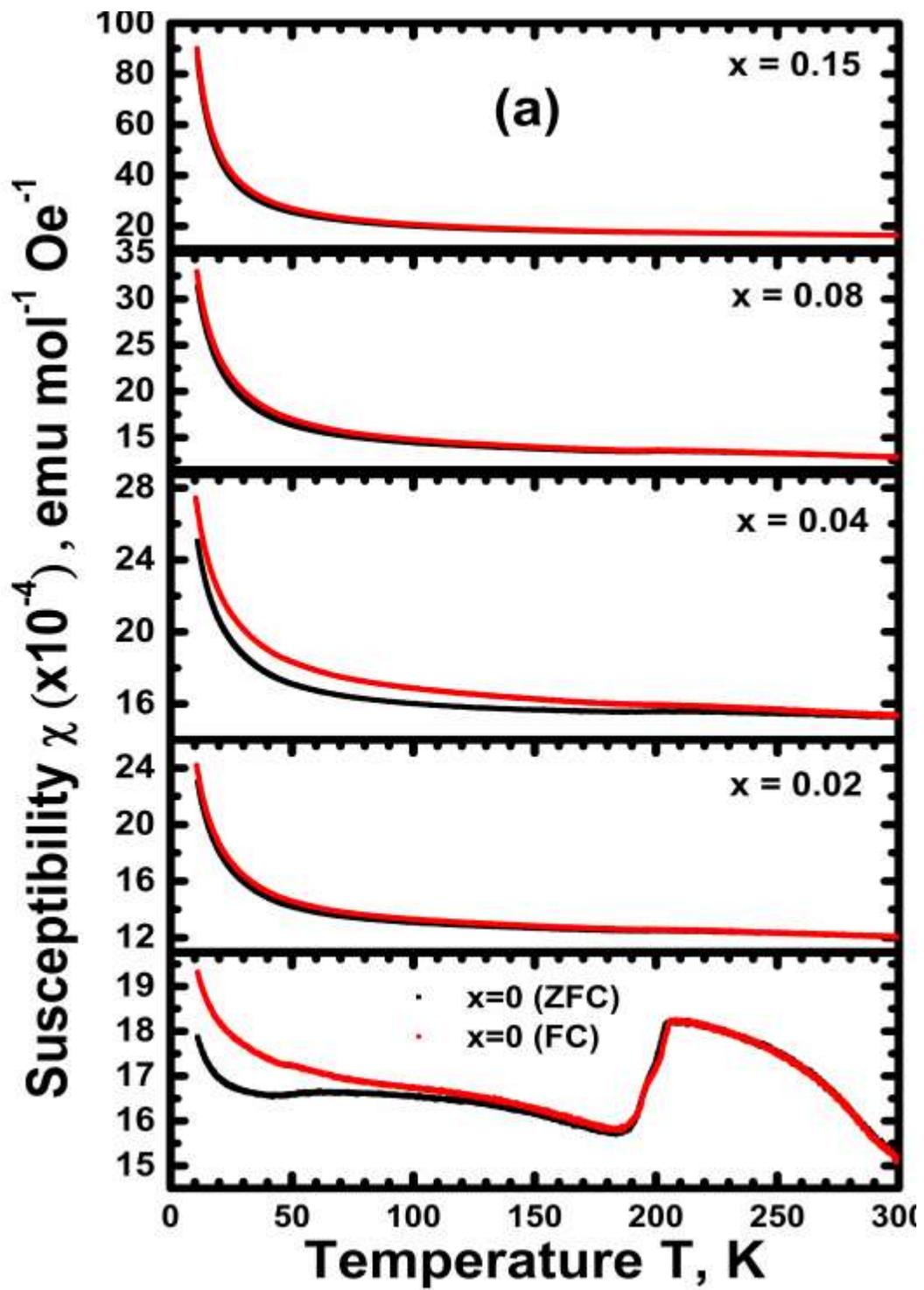

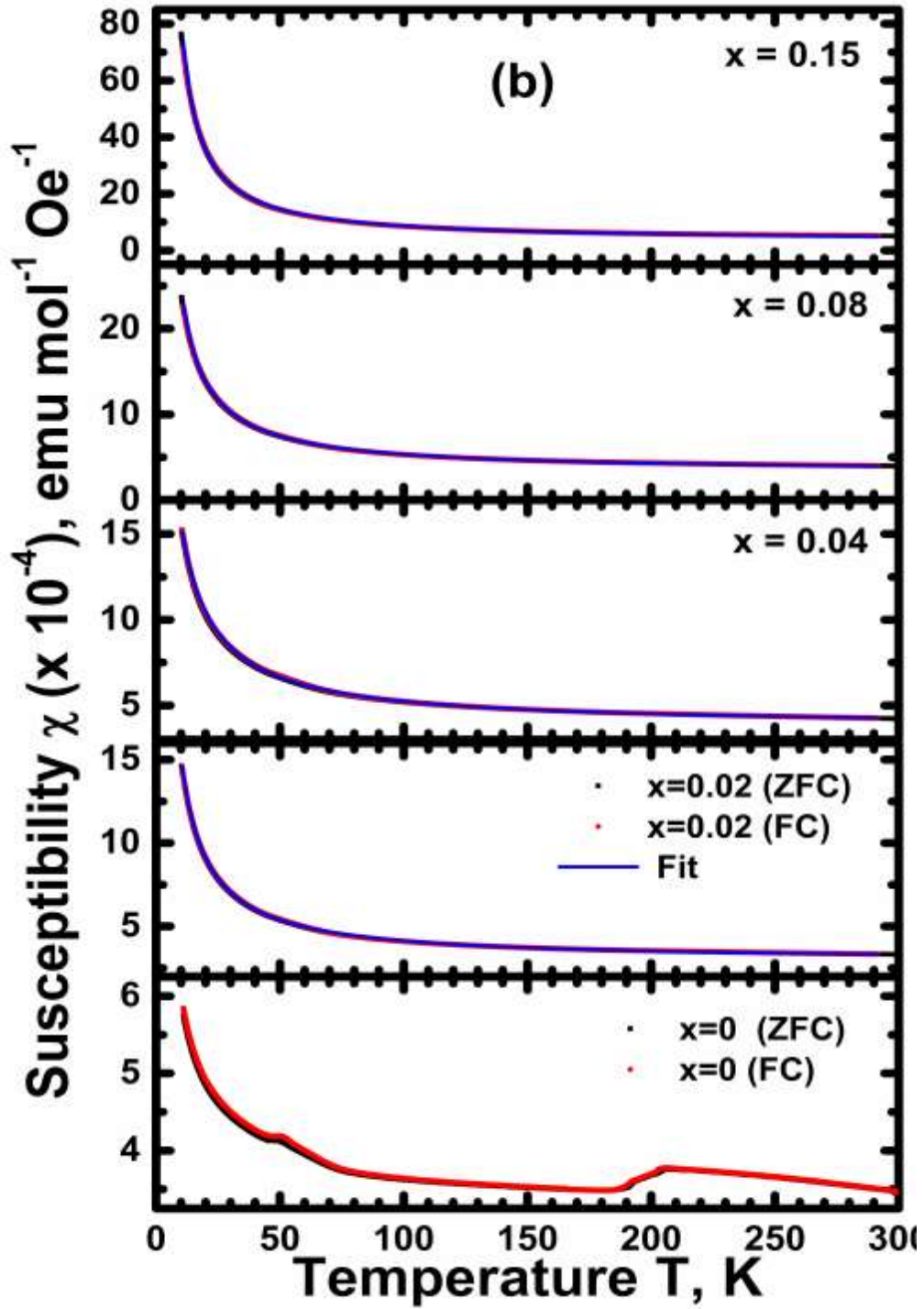

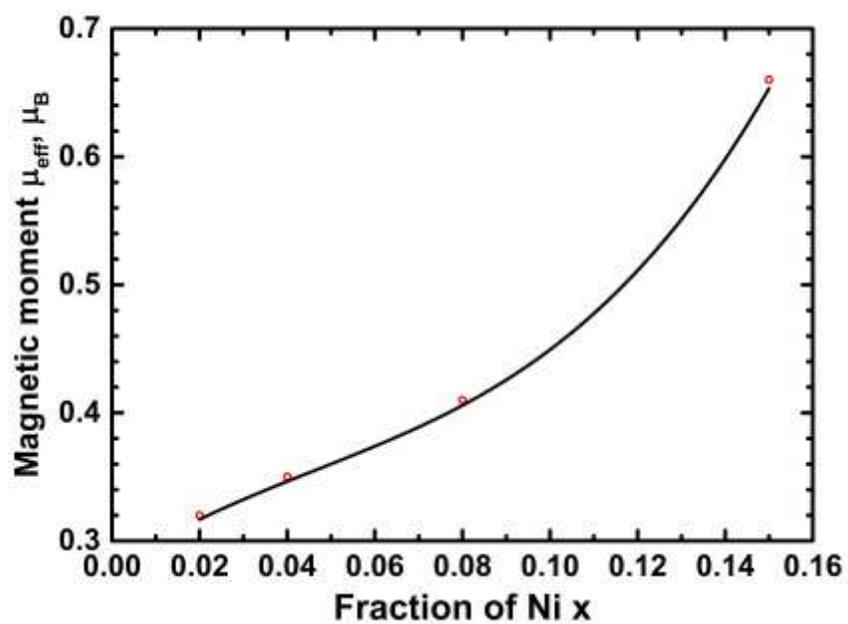
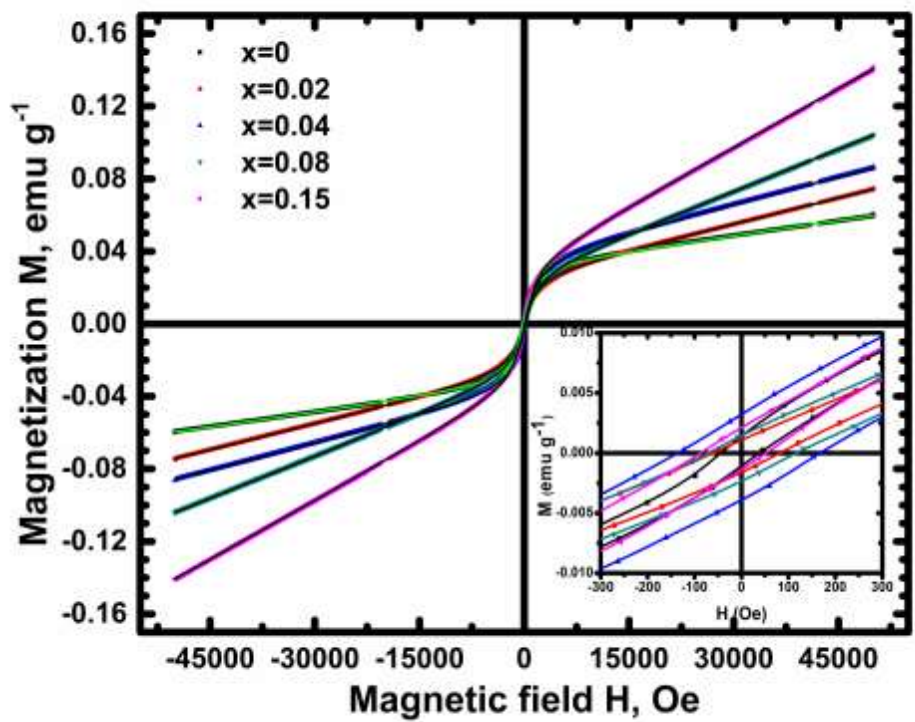